\documentclass[english,a4paper,12pt]{article}

\usepackage[T1]{fontenc}
\usepackage[latin9]{inputenc}
\usepackage{geometry}
\usepackage{amssymb}
\usepackage{graphicx}
\usepackage{cite}

\newcommand{\rmd}{\mathrm{d}}

\title{Nonlinear GARCH model and $1/f$ noise}

\author{A.~Kononovicius and J.~Ruseckas}
\date{}

\begin{document}

\maketitle

\begin{abstract}
Auto-regressive conditionally heteroskedastic (ARCH) family models
are still used, by practitioners in business and economic policy making,
as a conditional volatility forecasting models. Furthermore ARCH models
still are attracting an interest of the researchers. In this contribution
we consider the well known GARCH(1,1) process and its nonlinear modifications,
reminiscent of NGARCH model. We investigate the possibility to reproduce
power law statistics, probability density function and power spectral
density, using ARCH family models. For this purpose we derive stochastic
differential equations from the GARCH processes in consideration.
We find the obtained equations to be similar to a general class of
stochastic differential equations known to reproduce power law statistics.
We show that linear GARCH(1,1) process has power law distribution,
but its power spectral density is Brownian noise-like. However, the
nonlinear modifications exhibit both power law distribution and power
spectral density of the $1/f^{\beta}$ form, including $1/f$ noise.
\end{abstract}

\section{Introduction}

Forecasting volatility opens up a possibility to make better-informed
decisions. Thus this problem is of high interest to the practitioners
in business and economic policy making as well as to the scientific
community. The auto-regressive conditionally heteroskedastic (ARCH)
model and its generalization, known as GARCH, were proposed exactly
for this purpose \cite{Engle1982Econometrica,Bollerslev1986Econometrics}.
Since then numerous modifications of the seminal models proposed by
Engle and Bollerslev were introduced to serve varying purposes: from
financial market to macroeconomic modeling (see \cite{Bollerslev2008CREATES}
for a long list of the ARCH family models). Our particular interest,
in the context of this paper, lies in the continuous-time GARCH model
(COGARCH, see \cite{Nelson1990JEco,Lindner2008Springer,Kluppelberg2004JApplProbab,Kluppelberg2010SSRN})
and nonlinear variations of GARCH, such as NGARCH \cite{Engle1986EcoRev,Higgins1992IntEcoRev}
or MARCH \cite{Friedman1989Brookings}.

In order to evaluate the suitability of a given model one should compare
the features of the time series produced by the model with the actual
empirical data. It is known that high frequency time series of financial
data exhibit some universal statistical properties. Vast amounts of
historical stock price data around the world have helped to establish
a variety of so-called stylized facts \cite{Mantegna1995,Engle1998,Plerou1999,Engle2000,Ivanov2004,Bouchaud2004,Cont2005FIE,Cajueiro2009CSF,Podobnik2009PNAS,Ludescher2011EPL}
corresponding to the statistical signatures of financial processes.
One of the stylized facts concerns with autocorrelation function,
or, equivalently, power spectral density (PSD) of the time series.
There is empirical evidence that trading activity, trading volume,
and volatility are stochastic variables with the long-range correlation
\cite{Engle2001,Plerou2001,Gabaix2003} leading to $1/f$ type PSD.
Successful models should reproduce as many stylized facts as possible.
However, the $1/f$ type PSD is not accounted for in some widely used
models, such as ARCH family models. Therefore, it would be useful
to examine under which conditions power law PSD may be observed in
GARCH(1,1) and in nonlinear modifications of GARCH(1,1) models.

Power law statistics and especially $1/f$ noise are rather ubiquitous
phenomena observed in many different fields of science ranging from
natural phenomena to computer networks and financial markets \cite{Borland2012QFin,Chakraborti2011RQUF1,Gabaix2009AR,Karsai2012NIH,Parisi2013PhysRevE,Shao2011EPL}.
Since the discovery of $1/f$ noise numerous models and theories have
been proposed, for a recent review see \cite{Balandin2013}. A class
of nonlinear stochastic differential equations (SDEs) exhibiting power
law probability density function (PDF) and power law PSD in a wide
region of frequencies has been derived in \cite{Kaulakys2004PhysRevE,Kaulakys2006PhysA,Kaulakys2009JStatMech,Ruseckas2010PhysRevE}
starting from the point process model. Such nonlinear SDEs have been
used to describe signals in socio-economical systems \cite{Gontis2010PhysA,Mathiesen2013}.

Usually ARCH family models are calibrated by retro-fitting historical
data and thus may replicate patterns observed in the past \cite{Engle2008Wiley}.
Yet other approaches are also possible. For example, one may consider
comparing them to successful models from other frameworks. In this
paper we compare GARCH(1,1) and its nonlinear modifications to nonlinear
stochastic differential equations generating signals with power law
PSD.

This work is organized as follows: in Section \ref{sec:Stochastic-differential-equation}
we briefly introduce a class of nonlinear SDEs reproducing $1/f$
noise; in Section \ref{sec:GARCH(1,1)} we show that power law distributions
with varying exponents may be obtained from GARCH(1,1) process; in
Section \ref{sec:NGARCH} we consider a nonlinear modifications of
GARCH(1,1) process using which we reproduce $1/f$ noise. Section
\ref{sec:Discussion} summarizes our work.

\section{Stochastic differential equations generating signals with $1/f$
noise\label{sec:Stochastic-differential-equation}}

The nonlinear SDEs generating signals with power law steady state
PDF and $1/f^{\beta}$ PSD have been previously derived in Refs.~\cite{Kaulakys2004PhysRevE,Kaulakys2006PhysA,Kaulakys2009JStatMech}.
The general expression for the proposed class of It\^o SDEs is given
by
\begin{equation}
\rmd x=\sigma^{2}\left(\eta-\frac{1}{2}\lambda\right)x^{2\eta-1}\rmd t+\sigma x^{\eta}\rmd W_{t}\,.\label{eq:sde-ito}
\end{equation}
In the above $x$ is the signal, $\eta\neq1$ is the exponent of a
power law multiplicative noise, $\lambda$ defines the exponent of
the power law steady state PDF of the signal, while $W_{t}$ is the
standard Wiener process (one dimensional Brownian motion) and $\sigma$
is a scaling constant determining the intensity of noise. The nonlinear
SDE~(\ref{eq:sde-ito}) assumes the simplest form of the multiplicative
noise term, $\sigma x^{\eta}\rmd W_{t}$, although it may take other
forms as long as $\eta$ is the largest power of $x$ for large values
of $x$ \cite{Kaulakys2009JStatMech,Gontis2010PhysA}. Such nonlinear
SDEs have been used to describe signals in socio-economical systems
\cite{Gontis2010PhysA,Mathiesen2013}. In Refs.~\cite{Ruseckas2011epl,Kononovicius2012}
a nonlinear SDE similar to the one given by Eq.~(\ref{eq:sde-ito})
was derived by starting from a simple agent-based herding model, thus
providing agent-based reasoning for this class of SDEs.

The steady state PDF of Eq.~(\ref{eq:sde-ito}) has a power law form
$p(x)\sim x^{-\lambda}$ with the exponent $\lambda$. If $\lambda>1$
then $p(x)$ diverges as $x\rightarrow0$, therefore the diffusion
of the stochastic variable $x$ should be restricted from the side
of small values. This can be achieved by modifying Eq.~(\ref{eq:sde-ito}).
The simplest restriction of the diffusion is produced by the reflective
boundary conditions at the minimum value $x=x_{\mathrm{min}}$ and
the maximum value $x=x_{\mathrm{max}}$. Alternatively, one can modify
Eq.~(\ref{eq:sde-ito}) to get rapidly decreasing steady state PDF
when the stochastic variable $x$ acquires values outside of the interval
$[x_{\mathrm{min}},x_{\mathrm{max}}]$. For example, the steady state
PDF
\begin{equation}
p(x)\sim\frac{1}{x^{\lambda}}\exp\left(-\frac{x_{\mathrm{min}}^{m}}{x^{m}}-\frac{x^{m}}{x_{\mathrm{max}}^{m}}\right)
\end{equation}
with $m>0$ has a power law form inside of the interval $x_{\mathrm{min}}\ll x\ll x_{\mathrm{max}}$
and exponential cut-offs are present outside of this interval. Exponentially
restricted diffusion is generated by the SDE
\begin{equation}
\rmd x=\sigma^{2}\left[\eta-\frac{1}{2}\lambda+\frac{m}{2}\left(\frac{x_{\mathrm{min}}^{m}}{x^{m}}-\frac{x^{m}}{x_{\mathrm{max}}^{m}}\right)\right]x^{2\eta-1}\rmd t+\sigma x^{\eta}\rmd W_{t}\label{eq:4}
\end{equation}
which differs from Eq.~(\ref{eq:sde-ito}) only by a couple of additional
terms in the drift part of SDE.

One can estimate the PSD of the signals generated by the SDE~(\ref{eq:sde-ito})
by using the approximate scaling properties of the signals \cite{Ruseckas2014}.
The Wiener process scales as $\rmd W_{at}=a^{1/2}\rmd W_{t}$, thus
by changing the variable $x$ in Eq.~(\ref{eq:sde-ito}) to a scaled
variable $x_{s}=ax$ or by introducing the scaled time $t_{s}=a^{2(\eta-1)}t$
one obtains exactly the same SDE. This feature indicates that the
change of the scale of the stochastic variable $x$ and the change
of the time scale are statistically equivalent. Using the transition
probability (the conditional probability that at time $t$ the signal
has value $x'$ with the condition that at time $t=0$ the signal
had the value $x$) this equivalence may be mathematically expressed
as
\begin{equation}
aP(ax',t|ax,0)=P(x',a^{\mu}t|x,0)\,,\label{eq:trans-scaling}
\end{equation}
with the exponent $\mu$ being equal to $2(\eta-1)$. The discussed
scaling property Eq.~(\ref{eq:trans-scaling}) as well as power law
form of the steady state PDF $p(x)\sim x^{-\lambda}$ lead to the
PSD with power law behavior $S(f)\sim f^{-\beta}$, which is observed
in a wide range of frequencies. The power law exponent of the PSD
is given by \cite{Ruseckas2014}
\begin{equation}
\beta=1+\frac{\lambda-3}{2(\eta-1)}\,.\label{eq:beta}
\end{equation}

The restrictions imposed on diffusion at $x=x_{\mathrm{min}}$ and
$x=x_{\mathrm{max}}$ makes the scaling relationship Eq.~(\ref{eq:trans-scaling})
only approximate. This limits the power law part of the PSD to a certain
finite range of frequencies $f_{\mathrm{min}}\ll f\ll f_{\mathrm{max}}$.
Note, that power law behavior $1/f^{\beta}$ of the PSD for all frequencies
is physically impossible, because the total power of the signal then
would be infinite. Therefore, it is natural to consider signals with
the power law PSD in a limited range of frequencies. The frequency
range for the PSD of the signal generated by solving SDE~(\ref{eq:trans-scaling})
was estimated in Ref.~\cite{Ruseckas2014} as
\begin{eqnarray}
\sigma^{2}x_{\mathrm{min}}^{2(\eta-1)} & \ll & 2\pi f\ll\sigma^{2}x_{\mathrm{max}}^{2(\eta-1)},\qquad\eta>1,\label{eq:approx-range}\\
\sigma^{2}x_{\mathrm{max}}^{-2(1-\eta)} & \ll & 2\pi f\ll\sigma^{2}x_{\mathrm{min}}^{-2(1-\eta)},\qquad\eta<1.\nonumber 
\end{eqnarray}
From Eq.~(\ref{eq:approx-range}) it is evident that the width of
the frequency range can be increased by increasing the ratio between
the minimum and the maximum diffusion restriction boundary positions
$x_{\mathrm{max}}/x_{\mathrm{min}}$. In addition, the width also
depends on the multiplicative noise exponent $\eta$. Namely, the
width is zero if $\eta=1$ and increases with increasing $|\eta-1|$
\cite{Ruseckas2010PhysRevE}.

\begin{figure}
\centering{}\includegraphics[width=0.4\textwidth]{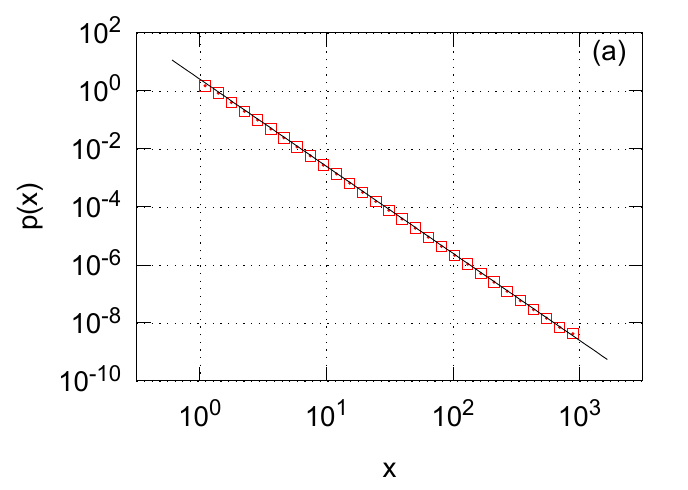}\hspace{0.1\textwidth}\includegraphics[width=0.4\textwidth]{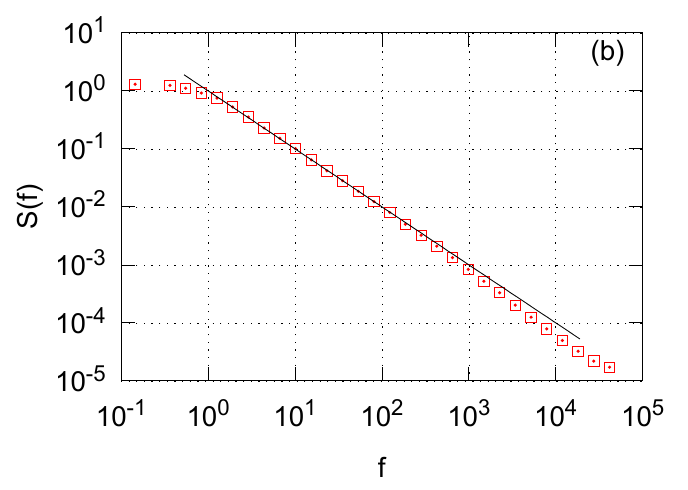}\caption{Statistical properties, PDF (a) and PSD (b), obtained by numerically
solving SDE~(\ref{eq:sde-ito}) (red squares). In the numerical computation
we have used the reflective boundaries placed at $x_{\mathrm{min}}$
and $x_{\mathrm{max}}$. Black curves show power law approximations
(a) $x^{-3}$ for numerically obtained PDF and (b) $1/f$ for the
PSD. The following parameter values were used: $\eta=2$, $\lambda=3$,
$x_{\mathrm{min}}=1$, $x_{\mathrm{max}}=10^{3}$, $\sigma=1$.\label{fig:1}}
\end{figure}

With $\lambda=3$, from Eq.~(\ref{eq:beta}) we obtain $\beta=1$
and SDE~(\ref{eq:sde-ito}) should generate a signal exhibiting $1/f$
noise. This case with $\eta=2$ is shown in Fig.~\ref{fig:1}. In
numerical computation we have used a modified Euler-Maruyama approximation
(the original Euler-Maruyama approximation is described in \cite{Kloeden1999Springer})
with a variable time step which decreases with the larger values of
$x$, as is described in \cite{Kaulakys2004PhysRevE,Kaulakys2006PhysA}.
We find a good agreement of the numerical results with analytical
predictions. Though the range of frequencies with $S(f)\sim1/f$ is
much narrower than expected, $1\ll f\ll10^{3}$ versus $1\ll f\ll10^{6}$
predicted by Eq.~(\ref{eq:approx-range}). This discrepancy arises
because Eq.~(\ref{eq:approx-range}) is only a qualitative estimation.
To obtain a more precise values of limiting frequencies one needs
to use more precise scaling properties of the nonlinear SDE.

\section{GARCH(1,1) process and stochastic differential equations\label{sec:GARCH(1,1)}}

Almost half a century ago Mandelbrot, Fama and others proposed an
idea that volatility fluctuations might be responsible for the fluctuating
nature of price change (return) dynamics \cite{Mandelbrot1963JBus,Fama1965JBus,Jeanblanc2009Springer}.
Actually intermittency of return time series is usually associated
with localized burst in volatility and thus frequently referred to
as volatility clustering \cite{Karsai2012NIH,Lo1991Econometrica,Ding1993JEmpFin,Gontis2012ACS}.
It was proven that modeling temporal dynamics of second-order moment,
known as heteroskedasticity \cite{Engle1982Econometrica}, may allow
better performing option-price models \cite{Heston1993RevFinStud,Potters1998EPL,Fouque2000CUP}.

In his seminal article \cite{Engle1982Econometrica} R.~F.~Engle
laid foundations to the autoregressive conditional heteroskedasticity
(ARCH) models by proposing to split a heteroskedastic variable $z$
(e.g., return) into a stochastic part $\omega_{t}$ and time dependent
volatility (standard deviation) $\sigma_{t}$,

\begin{equation}
z_{t}=\sigma_{t}\omega_{t}\,.\label{eq:z}
\end{equation}
Stochastic part $\omega_{t}$ may follow any distribution, but a common
choice is the Gaussian distribution, though other distributions such
as the $q$-Gaussian distribution might be also considered \cite{Borland2002PRL,Gontis2014PLOS}.
In this paper we will use the Gaussian distribution with zero mean
$\langle\omega_{t}\rangle=0$ and unit variance $\langle\omega_{t}^{2}\rangle=1$.
As the process modeling the evolution of the standard deviation time
series $\sigma_{t}$ we choose GARCH(1,1) process, which is defined
as an iterative equation of the following form:
\begin{equation}
\sigma_{t}^{2}=a+bz_{t-1}^{2}+c\sigma_{t-1}^{2}\,.\label{eq:garch}
\end{equation}
Using Eq.~(\ref{eq:z}) the GARCH(1,1) process can be written as
\begin{equation}
\sigma_{t}^{2}=a+b\sigma_{t-1}^{2}\omega_{t-1}^{2}+c\sigma_{t-1}^{2}\,.
\end{equation}

GARCH(1,1) process can be approximated as a continuous time SDE by
considering the diffusion limit of this process \cite{Nelson1990JEco,Lindner2008Springer,Kluppelberg2004JApplProbab,Kluppelberg2010SSRN}.
Usually the parameters of the GARCH process are obtained by retro-fitting
empirical data and thus the actual values of the parameters $a$,
$b$ and $c$ are tied to a particular discretization time step. Taking
this into account let us rewrite the above as \cite{Lindner2008Springer}

\begin{eqnarray}
\sigma_{kh,h}^{2} & = & a_{h}+b_{h}\sigma_{(k-1)h,h}^{2}\omega_{(k-1)h,h}^{2}+c_{h}\sigma_{(k-1)h,h}^{2}\nonumber \\
 & = & \sigma_{(k-1)h,h}^{2}+a_{h}-(1-b_{h}-c_{h})\sigma_{(k-1)h,h}^{2}+b_{h}\sigma_{(k-1)h,h}^{2}\left(\omega_{(k-1)h,h}^{2}-1\right)\,.\label{eq:garch-expanded}
\end{eqnarray}
Here $h$ is a time series discretization period ($t=kh$, where $k=1,2,3,\ldots$),
while subscripts indicate that the parameters depend on $h$. Assuming
that $h$ is infinitesimally small, $h\rightarrow0$ we introduce
the continuous time parameters $A$, $B$ and $C$ related to the
parameters $a_{h}$, $b_{h}$ and $c_{h}$ by the equations \cite{Lindner2008Springer}
\begin{equation}
a_{h}=Ah,\quad1-b_{h}-c_{h}=Ch,\quad2b_{h}^{2}=B^{2}h\,.\label{eq:cont-time-params}
\end{equation}
Note, that the parameters $A$, $B$, $C$ do not depend on $h$.
Using Eq.~(\ref{eq:cont-time-params}) we can rewrite Eq.~(\ref{eq:garch-expanded})
as 
\begin{equation}
\sigma_{kh,h}^{2}=\sigma_{(k-1)h,h}^{2}+\left(A-C\sigma_{(k-1)h,h}^{2}\right)h+|B|\sqrt{\frac{h}{2}}\sigma_{(k-1)h,h}^{2}\left(\omega_{(k-1)h,h}^{2}-1\right)\,.
\end{equation}
The stochastic variable $\omega_{(k-1)h,h}^{2}-1$ we approximate
as a normally distributed stochastic variable with zero mean and variance
equal to $2$, since $\langle\omega^{2}-1\rangle=0$ and $\langle(\omega^{2}-1)^{2}\rangle=k_{\omega}-1=2$.
This gives
\begin{equation}
\sigma_{kh,h}^{2}=\sigma_{(k-1)h,h}^{2}+\left(A-C\sigma_{(k-1)h,h}^{2}\right)h+|B|\sigma_{(k-1)h,h}^{2}\sqrt{h}\varepsilon_{k}\,,\label{eq:linear-diff}
\end{equation}
where $\varepsilon_{k}$ is a normally distributed stochastic variable
with zero mean and unit variance. Note, that the last equation has
the form of a difference equation, similar to the Euler-Marujama approximation
used to numerically solve SDEs. Thus by taking the small time step
limit $h\rightarrow0$ one can obtain the SDE for the variable $y_{kh,h}=\sigma_{kh,h}^{2}$
from Eq.~(\ref{eq:linear-diff}):
\begin{equation}
\rmd y=(A-Cy)\rmd t+|B|y\rmd W_{t}\,.
\end{equation}
This equation can be written in the form
\begin{equation}
\rmd y=B^{2}\left(1-\frac{1}{2}\lambda+\frac{1}{2}\frac{y_{\mathrm{min}}}{y}\right)y\rmd t+|B|y\rmd W_{t}\,,\label{eq:sde-garch}
\end{equation}
where

\begin{eqnarray}
\lambda & = & 2+\frac{2C}{B^{2}}=2+\frac{1-b_{h}-c_{h}}{b_{h}^{2}}\,,\qquad y_{\mathrm{min}}=\frac{2A}{B^{2}}=\frac{a}{b_{h}^{2}}\,.
\end{eqnarray}
The SDE~(\ref{eq:sde-garch}) is a special case of the SDE~(\ref{eq:sde-ito})
with $\eta=1$. As is discussed in Section \ref{sec:Stochastic-differential-equation},
steady state PDF $p(y)$ thus should have power law tail with the
exponent $\lambda$, $p(y)\sim y^{-\lambda}$. Analytical prediction
Eq.~(\ref{eq:beta}) for the power law exponent in the PSD diverges,
but it is worth to note that the SDE~(\ref{eq:sde-garch}) is similar
to geometric Brownian motion, PSD of which has the power law exponent
$\beta=2$.

\begin{figure}
\begin{centering}
\includegraphics[width=0.4\textwidth]{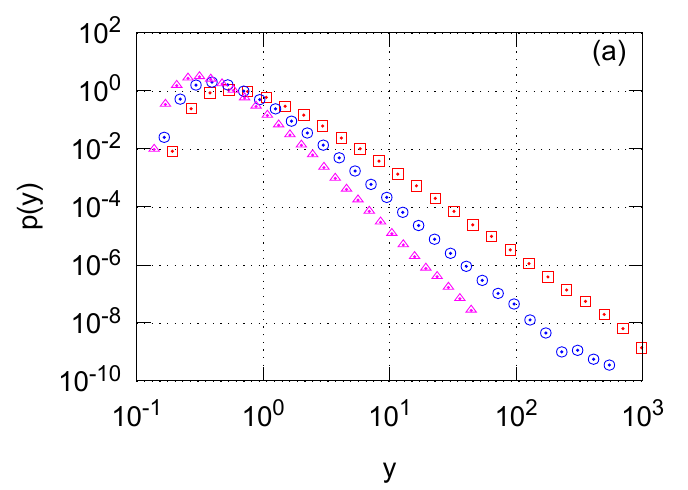}\hspace{0.1\textwidth}\includegraphics[width=0.4\textwidth]{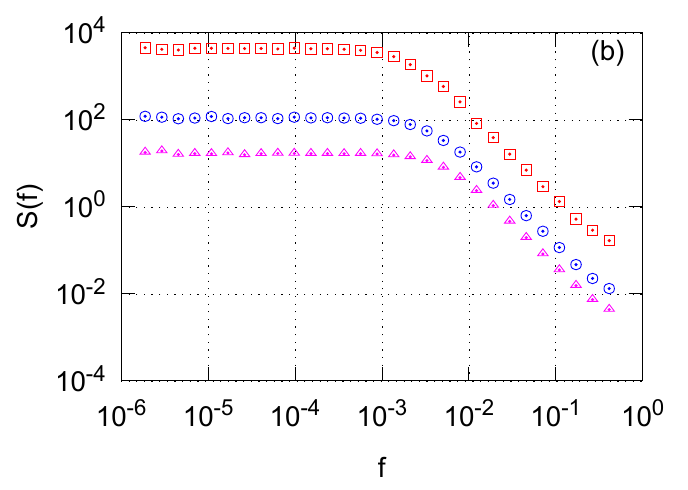}
\par\end{centering}

\caption{Statistical properties, PDF (a) and PSD (b), of numerically evaluated
linear GARCH(1,1) process Eq.~(\ref{eq:garch}). The GARCH(1,1) process
parameters were set as follows: $a=0.015$, $b=0.1$, $c=0.89$ (red
squares), $0.88$ (blue circles) and $0.87$ (magenta triangles).\label{fig:linear}}
\end{figure}

The PDF and the PSD of the time series numerically obtained using
GARCH(1,1) process, Eq.~(\ref{eq:garch}), are shown in Fig.~\ref{fig:linear}.
The analytical predictions of the power law exponents in the PDF and
the PSD are in good agreement with the numerical results.

\section{Nonlinear GARCH(1,1) process generating signals with $1/f$ noise\label{sec:NGARCH}}

The mathematical form of Eq.~(\ref{eq:beta}) suggests that it is
possible to obtain other values of the power law exponent $\beta$
as long as $\eta\neq1$. In our previous work we have shown that $\eta>1$
cases work very well for the modeling of high-frequency trading activity
as well as high-frequency absolute returns of the financial markets
\cite{Gontis2008PhysA,Gontis2010PhysA,Gontis2010Intech}, although
theoretically $\eta<1$ is also possible \cite{Ruseckas2010PhysRevE}.
One can obtain the $\eta>1$ case by considering the following modifications
of GARCH(1,1) process:
\begin{eqnarray}
\sigma_{t}^{2} & = & a+b\sigma_{t-1}^{\mu}\omega_{t-1}^{\mu}+c\sigma_{t-1}^{2}\,,\label{eq:ngarch-2-1}
\end{eqnarray}
where $\mu>2$ is an odd integer, and
\begin{equation}
\sigma_{t}^{2}=a+b\sigma_{t-1}^{\mu}|\omega_{t-1}|^{\mu}+\sigma_{t-1}^{2}-c\sigma_{t-1}^{\mu}\,,\label{eq:ngarch-3-1}
\end{equation}
where $\mu$ may be any positive real number. Nonlinear GARCH model
of a similar form was considered by Engle and Bollerslev in \cite{Engle1986EcoRev}.
Nonlinear model proposed by Engle and Bollerslev had a form of Eq.~(\ref{eq:ngarch-2-1}),
but absolute value of $\omega_{t-1}$ was taken prior to raising it
to a generalized power $\mu$. Engle and Bollerslev found that $\mu\lessapprox2$
for most empirical timeseries they have considered. Another take at
nonlinear GARCH model can be found in work by Higgins and Bera \cite{Higgins1992IntEcoRev},
although they considered dynamics not of the standard deviation (as
we do), but of the higher order moment, $\sigma_{t}^{\mu}$. Note,
that the last two terms in Eq.~(\ref{eq:ngarch-3-1}) can be seen
as the first two terms in the power series expansion of a more general
function of standard deviation $f(\sigma_{t})$. In contrast to Eq.~(\ref{eq:garch})
for the GARCH(1,1) process, Eqs.~(\ref{eq:ngarch-2-1}) and (\ref{eq:ngarch-3-1})
do not ensure the positivity of $\sigma_{t}^{2}$. In order to avoid
negative values of $\sigma_{t}^{2}$ we consider Eqs.~(\ref{eq:ngarch-2-1})
and (\ref{eq:ngarch-3-1}) together with a reflective boundary at
$\sigma_{t}^{2}=0$.

Let us first consider the diffusion limit of Eq.~(\ref{eq:ngarch-2-1}).
We proceed similarly as in Section~\ref{sec:GARCH(1,1)}. Taking
into account the relation to physical time via time series sampling,
this iterative equation may be rewritten as follows:
\begin{equation}
\sigma_{kh,h}^{2}=\sigma_{(k-1)h,h}^{2}+a_{h}-(1-c_{h})\sigma_{(k-1)h,h}^{2}+b_{h}\sigma_{(k-1)h,h}^{\mu}\omega_{(k-1)h,h}^{\mu}\,.\label{eq:ngarch-2-2}
\end{equation}
Assuming that the time discretization period $h$ is infinitesimally
small, $h\rightarrow0$, we can introduce the coefficients $A$, $B$
and $C$ via the equations
\begin{equation}
a_{h}=Ah\,,\qquad1-c_{h}=Ch\,,\qquad\langle\omega^{2\mu}\rangle b_{h}^{2}=B^{2}h
\end{equation}
and rewrite Eq.~(\ref{eq:ngarch-2-2}) as
\begin{equation}
\sigma_{kh,h}^{2}=\sigma_{(k-1)h,h}^{2}+\left(A-C\sigma_{(k-1)h,h}^{2}\right)h+|B|\sqrt{\frac{h}{\langle\omega^{2\mu}\rangle}}\sigma_{(k-1)h,h}^{\mu}\omega_{(k-1)h,h}^{\mu}\,.
\end{equation}
We approximate the stochastic variable $\omega_{(k-1)h,h}^{\mu}$
as a normal stochastic variable with zero mean (given that $\mu$
is odd) and variance $\langle\omega^{2\mu}\rangle$. This gives
\begin{equation}
\sigma_{kh,h}^{2}=\sigma_{(k-1)h,h}^{2}+\left(A-C\sigma_{(k-1)h,h}^{2}\right)h+|B|\sqrt{h}\sigma_{(k-1)h,h}^{\mu}\varepsilon_{k}\,,\label{eq:nlin-diff}
\end{equation}
where $\varepsilon_{k}$ is a normally distributed stochastic variable
with zero mean and unit variance. By taking the small time step limit
$h\rightarrow0$, one can obtain the SDE for the variable $y_{kh,h}=\sigma_{kh,h}^{2}$
from Eq.~(\ref{eq:nlin-diff}):
\begin{equation}
\rmd y=\left(\frac{A}{y^{\mu-1}}-\frac{C}{y^{\mu-2}}\right)y^{\mu-1}\rmd t+|B|y^{\frac{\mu}{2}}\rmd W_{t}\,.\label{eq:sde-11}
\end{equation}
By introducing the parameters
\begin{eqnarray}
y^{(1)} & = & \left(\frac{2A}{(\mu-1)B^{2}}\right)^{\frac{1}{\mu-1}}\,,\qquad y^{(2)}=\left(\frac{2C}{(\mu-2)B^{2}}\right)^{\frac{1}{\mu-2}}
\end{eqnarray}
Eq.~(\ref{eq:sde-11}) can be rewritten as
\begin{equation}
\rmd y=B^{2}\left[\frac{1}{2}(\mu-1)\left(\frac{y^{(1)}}{y}\right)^{\mu-1}-\frac{1}{2}(\mu-2)\left(\frac{y^{(2)}}{y}\right)^{\mu-2}\right]y^{\mu-1}\rmd t+|B|y^{\frac{\mu}{2}}\rmd W_{t}\,.\label{eq:13-1}
\end{equation}
From the Fokker-Planck equation corresponding to Eq.~(\ref{eq:13-1})
one can obtain the steady state PDF $p(y)$ of the signal generated
by Eq.~(\ref{eq:13-1}). The steady state PDF $p(y)$ has a power
law tail with the exponent $\lambda=\mu$, while $y^{(1)}$ and $y^{(2)}$
shape the exponential cutoff:

\begin{equation}
p(y)\sim\frac{1}{y^{\mu}}\exp\left[-\left(\frac{y^{(1)}}{y}\right)^{\mu-1}+\left(\frac{y^{(2)}}{y}\right)^{\mu-2}\right]\,.
\end{equation}
Eq.~(\ref{eq:13-1}) has the general form of SDE~(\ref{eq:sde-ito})
with the parameters $\lambda=\mu$ and $\eta=\mu/2$. Thus the PSD
of $y_{t}$ time series should have a frequency range with the power
law behavior of the PSD $S(f)$ given by

\begin{equation}
S(f)\sim\frac{1}{f^{\beta}}\,,\qquad\beta=1+\frac{\mu-3}{\mu-2}\,.\label{eq:beta-mu}
\end{equation}
Note, that we get $1/f$ PSD when $\mu=3$.

\begin{figure}
\begin{centering}
\includegraphics[width=0.4\textwidth]{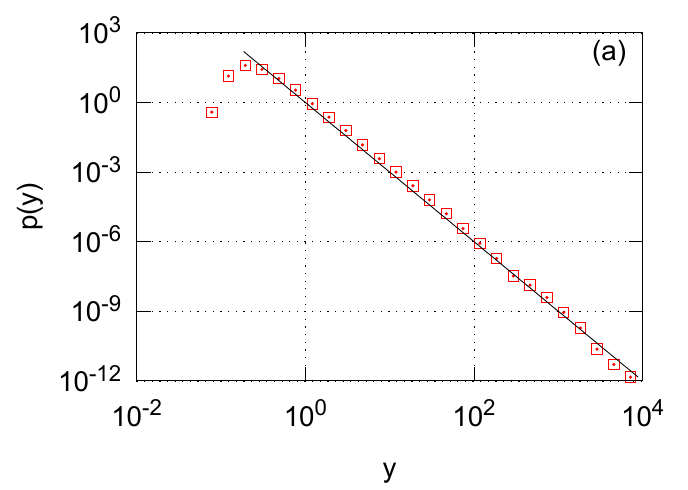}\hspace{0.1\textwidth}\includegraphics[width=0.4\textwidth]{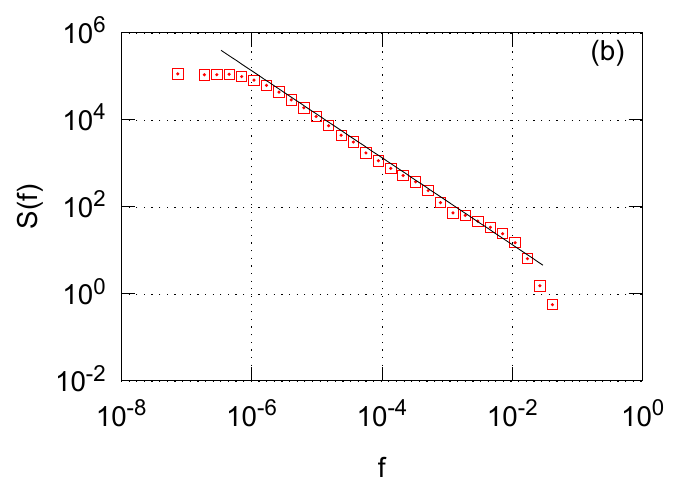}
\par\end{centering}

\caption{Statistical properties, PDF (a) and PSD (b), of numerically evaluated
nonlinear GARCH process Eq.~(\ref{eq:ngarch-2-1}) with $\mu=3$
(red squares). Black curves show power law approximations (a) $x^{-3}$
for numerically obtained PDF and (b) $1/f$ for the PSD. Other nonlinear
GARCH process parameters were set as follows: $a=10^{-6}$, $b=10^{-3}$,
$c=1$.\label{fig:nonlinear-1f}}
\end{figure}

The PDF and the PSD of the time series numerically obtained using
Eq.~(\ref{eq:ngarch-2-1}) with $\mu=3$ are shown in Fig.~\ref{fig:nonlinear-1f}.
In the numerical calculations we used reflective boundary at $\sigma_{t}=0$.
The analytical predictions of the power law exponents $\lambda=3$
in the PDF and $\beta=1$ in the PSD are in good agreement with the
numerical results. With the chosen parameters in Eq.~(\ref{eq:ngarch-2-1})
we are able to reproduce $1/f$ spectrum over almost $4$ decades
of frequency $f$, see Fig.~\ref{fig:nonlinear-1f}(b).

Now let us consider the diffusion limit of Eq.~(\ref{eq:ngarch-3-1}).
Eq.~(\ref{eq:ngarch-3-1}) may be rewritten by taking into account
the relation to physical time:
\begin{equation}
\sigma_{kh,h}^{2}=\sigma_{(k-1)h,h}^{2}+a_{h}-c_{h}\sigma_{(k-1)h,h}^{\mu}+b_{h}\sigma_{(k-1)h,h}^{\mu}|\omega_{(k-1)h,h}|^{\mu}\,.\label{eq:ngarch-3-2}
\end{equation}
Approximating $|\omega_{(k-1)h,h}|^{\mu}$ as a normal stochastic
variable with the mean $\bar{\omega}_{\mu}=\langle|\omega|^{\mu}\rangle$
and the variance $\hat{\omega}_{\mu}=\langle\left[|\omega|^{\mu}-\bar{\omega}_{\mu}\right]^{2}\rangle$
yields
\begin{equation}
\sigma_{kh,h}^{2}=\sigma_{(k-1)h,h}^{2}+a_{h}+\left(b_{h}\bar{\omega}_{\mu}-c_{h}\right)\sigma_{(k-1)h,h}^{\mu}+b_{h}\sigma_{(k-1)h,h}^{\mu}\sqrt{\hat{\omega}_{\mu}}\varepsilon_{k}\,.\label{eq:29}
\end{equation}
Here $\varepsilon_{k}$ is a normally distributed stochastic variable
with zero mean and unit variance. Assuming that the time discretization
period $h$ is infinitesimally small, $h\rightarrow0$, we can write
Eq.~(\ref{eq:29}) as
\begin{equation}
\sigma_{kh,h}^{2}=\sigma_{(k-1)h,h}^{2}+\left(A+C\sigma_{(k-1)h,h}^{\mu}\right)h+|B|\sqrt{h}\sigma_{(k-1)h,h}^{\mu}\varepsilon_{k}\,,
\end{equation}
where the coefficients $A$, $B$ and $C$ are introduced via the
equations
\begin{equation}
a_{h}=Ah\,,\qquad b_{h}\bar{\omega}_{\mu}-c_{h}=Ch\,,\qquad\hat{\omega}_{\mu}b_{h}^{2}=B^{2}h\,.\label{eq:params-3}
\end{equation}
Note, that depending on the values of the parameters $b_{h}$ and
$c_{h}$, the parameter $C$ can be negative as well as positive.
By taking the small time step limit $h\rightarrow0$ we can obtain
the SDE for the variable $y_{kh,h}=\sigma_{kh,h}^{2}$:
\begin{equation}
\rmd y=\left(\frac{A}{y^{\mu-1}}+\frac{C}{y^{\frac{\mu}{2}-1}}\right)y^{\mu-1}\rmd t+|B|y^{\frac{\mu}{2}}\rmd W_{t}\,.
\end{equation}
This equation can be written as
\begin{equation}
\rmd y=B^{2}\left[\frac{1}{2}(\mu-1)\left(\frac{y^{(1)}}{y}\right)^{\mu-1}+\frac{1}{2}\mathrm{sign}(C)\left(\frac{\mu}{2}-1\right)\left(\frac{y^{(3)}}{y}\right)^{\frac{\mu}{2}-1}\right]y^{\mu-1}\rmd t+|B|y^{\frac{\mu}{2}}\rmd W_{t}\,,\label{eq:13-2}
\end{equation}
where
\begin{eqnarray}
y^{(1)} & = & \left(\frac{2A}{(\mu-1)B^{2}}\right)^{\frac{1}{\mu-1}}\,,\qquad y^{(3)}=\left(\frac{4|C|}{\left(\mu-2\right)B^{2}}\right)^{\frac{2}{\mu-2}}\,.
\end{eqnarray}
The steady state PDF $p(y)$ of the signal generated by Eq.~(\ref{eq:13-2})
has a power law tail with the exponent $\lambda=\mu$, while $y^{(1)}$
and $y^{(3)}$ shape the exponential cutoff:

\begin{equation}
p(y)\sim\frac{1}{y^{\mu}}\exp\left[-\left(\frac{y^{(1)}}{y}\right)^{\mu-1}-\mathrm{sign}(C)\left(\frac{y^{(3)}}{y}\right)^{\mu-2}\right]\,.
\end{equation}
Eq.~(\ref{eq:13-2}) has the general form of SDE~(\ref{eq:sde-ito})
with the parameters $\lambda=\mu$ and $\eta=\mu/2$; the parameters
are the same as as in the previous case, Eq.~(\ref{eq:ngarch-2-1}).
Therefore, the PSD of $y_{t}$ time series should have a frequency
range where Eq.~(\ref{eq:beta-mu}) holds.

\begin{figure}
\begin{centering}
\includegraphics[width=0.4\textwidth]{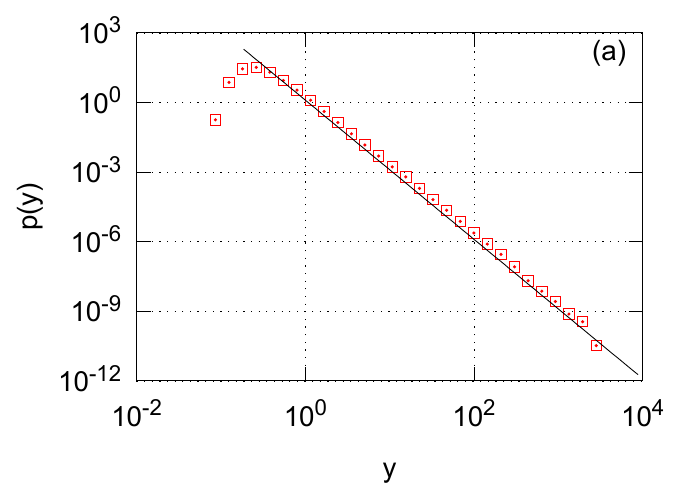}\hspace{0.1\textwidth}\includegraphics[width=0.4\textwidth]{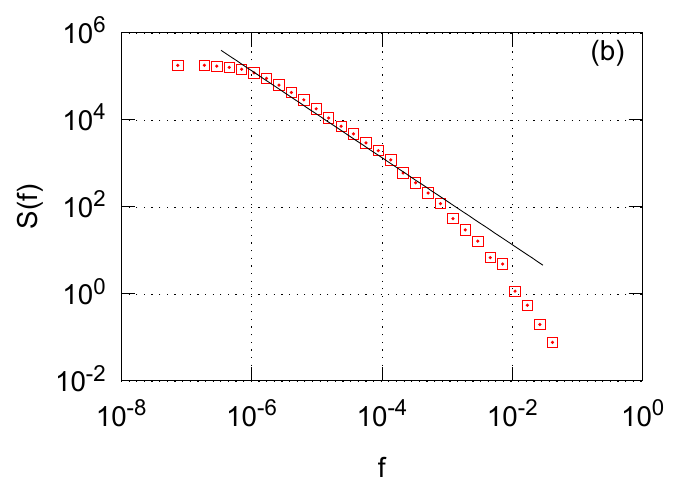}
\par\end{centering}

\caption{Statistical properties, PDF (a) and PSD (b), of numerically evaluated
nonlinear GARCH process Eq.~(\ref{eq:ngarch-3-1}) with $\mu=3$
(red squares). Black curves show power law approximations (a) $x^{-3}$
for numerically obtained PDF and (b) $1/f$ for the PSD. Other nonlinear
GARCH process parameters were set as follows: $a=10^{-6}$, $b=10^{-3}$,
$c=2\sqrt{\frac{2}{\pi}}\cdot10^{-3}\approx1.595769\cdot10^{-3}$.
The chosen values of the parameters $b$ and $c$ lead to $C\approx0$
and, consequently, $y^{(3)}\approx0$ in the SDE~(\ref{eq:13-2})
describing the diffusion limit of this nonlinear GARCh process.\label{fig:nonlinear-1f-2}}
\end{figure}

The PDF and the PSD of the time series numerically obtained using
Eq.~(\ref{eq:ngarch-3-1}) with $\mu=3$ are shown in Fig.~\ref{fig:nonlinear-1f-2}.
We have chosen the parameters $b$ and $c$ in Eq.~(\ref{eq:ngarch-3-1})
in such a way that the parameter $C$ given by Eq.~(\ref{eq:params-3})
becomes zero. In the numerical calculations we used reflective boundary
at $\sigma_{t}=0$ by setting $\sigma_{t}^{2}$ to zero when $\sigma_{t}^{2}$
becomes negative. The analytical predictions of the power law exponents
$\lambda=3$ in the PDF and $\beta=1$ in the PSD are in good agreement
with the numerical results. With the chosen parameters in Eq.~(\ref{eq:ngarch-3-1})
we are able to reproduce $1/f$ spectrum, but now over only $3$ decades
of frequency $f$, see Fig.~\ref{fig:nonlinear-1f-2}(b).

\section{Conclusions\label{sec:Discussion}}

In summary, we have proposed two possible nonlinear modifications of
a GARCH(1,1) process, Eqs.~(\ref{eq:ngarch-2-1}) and (\ref{eq:ngarch-3-1}).
Comparing the diffusion limit of the proposed nonlinear GARCH processes
with the known nonlinear SDE (\ref{eq:sde-ito}) generating signals
with $1/f^{\beta}$ PSD we obtain the conditions when the nonlinear
GARCH processes yield $1/f^{\beta}$ PSD too. Numerical evaluation
of Eqs.~(\ref{eq:ngarch-2-1}) and (\ref{eq:ngarch-3-1}) with suitably
chosen parameters confirms the presence of a wide power law region
in the PSD of the time series. As should have been expected, the linear
GARCH(1,1) process (\ref{eq:garch}) does not reproduce $1/f$ spectrum.
In addition to power law PSD, linear and nonlinear GARCH(1,1) processes
resulted in power law distributions.

The results obtained in the paper are especially interesting as $1/f$
noise is often linked to a concept of long-range memory, which is
considered to be one of the stylized facts of the financial markets
as well as other socio-economic systems. The obtained results and
proposed nonlinear GARCH processes should be useful for creation and
application of ARCH family models that correctly reproduce the PSD
of the financial time series as well.

%\bibliographystyle{elsarticle-num}
%\bibliography{garch}

\end{document}